# Dynamic Control of Momentum-Polarization Photoluminescence States with Liquid-Crystal-tuned Nanocavities

**Chengkun Dong, Matthew R. Chua, Rasna Maruthiyodan Veetil, T. Thu Ha Do, Lu Ding, Deepak K. Sharma, Jun Xia, Ramón Paniagua-Domínguez\***

C. Dong, M. R. Chua, R. M. Veetil, T. T. H. Do, L. Ding, Deepak K. Sharma, R. Paniagua-Dominguez
Institute of Materials Research and Engineering (IMRE), Agency for Science, Technology and Research (A*STAR), 2 Fusionopolis Way, Innovis #08-03, Singapore, 138634
Republic of Singapore
\*Email: ramon_paniagua@imre.a-star.edu.sg

C. Dong, J. Xia
School of Electronic Science and Engineering, Joint International Research Laboratory of Information Display and Visualization, Southeast University, Nanjing 210096, China

**Abstract**

Dynamic control of light, and in particular beam steering, is pivotal in various optical applications, including telecommunications, LiDAR, and biomedical imaging. Traditional approaches achieve this by interfacing a tunable modulating device with an external light source, facing challenges in achieving compact devices. Here, we introduce a dynamic photoluminescence (PL) modulating device, with which the properties of light directly emitted by a quasi-two-dimensional perovskite (in particular its directionality and polarization) can be modified continuously and over a large range. The device is based on a liquid-crystal-tunable Fabry-Perot (FP) nanocavity and uses the FP energy-momentum dispersion and spin-orbit coupling between the excitons and the cavity modes to enable this dynamic control over the emitted radiation. With this device, we achieve electrically-controlled, continuous and variable emission angles up to a maximum of 28°, as well as manipulation of the PL polarization state, enabling both the creation of polarization gradients and the achievement of polarization conversion at specific emission angles. Moreover, due to its resonant character, a 3-fold increase in the emission intensity is observed, as confirmed through time-resolved photoluminescence (TRPL) measurements. Our approach leverages the unique properties of actively tunable birefringent nanocavities to improve emission directivity, angle tunability and polarization control, presenting a promising solution for next-generation, deeply integrated beam steering devices.

**Introduction**

Beam steering technologies play a critical role in a variety of optical systems, ranging from telecommunications[1,2] and LiDAR[3,4] to biomedical imaging[5,6]. Recently, a novel approach for realizing beam steering devices has emerged using actively tunable metasurfaces and nanocavities[7-10]. These devices utilize various tunable media, such as liquid crystals (LC)[11-16], phase change materials[17-20], two-dimensional (2D) materials[21-24], transparent conducting oxides[25-27] and semiconductors[28,29], to modulate external reflected or transmitted light in an ultra-thin fashion, thereby enabling a reduction in size of these beam deflection devices. They, however, still require precisely controlled external illumination, limiting further integration. Also, in many cases, achieving large-angle and continuous beam steering remains a challenge[30,31].

Modifying the properties of light directly emitted by active/gain materials, in which one can control the far-field emission pattern and/or the polarization state, can be achieved by integrating these materials with nanophotonic cavities[32-35]. In this regard, a large number of studies have thus far demonstrated static control of the directionality and/or polarization of photoluminescence (PL) or lasing[36-44].

On the other hand, the number of studies focusing on the use of tunable nanophotonic environments for dynamic control of PL remains much lower (even more so, in the case of lasing[45-49]). In this regard, it has been recently shown that dynamic steering of incoherent emission can be achieved by imposing a transient spatially structured refractive index profile in the form of a blazed grating on a light-emitting semiconductor metasurface[50]. This index modulation, induced by a spatially structured optical pump, imparts additional momentum to the spontaneous emission from quantum dots, resulting in an angular shift of the far field emission. While having very interesting features, e. g. its ultra-fast character, this method relies on a significant number of bulky optical components to achieve dynamic tunability, posing major difficulties for miniaturization. LCs have also been used to dynamically manipulate photoluminescence by embedding emitters in tunable birefringent cavities. These systems enable electrical tuning of exciton-polariton resonances[51] and polarization[52]. In such anisotropic environments, spin–orbit coupling effects emerge, and broken symmetry can lead to Rashba-Dresselhaus spin splitting.

In this work, we present a dynamic PL beam steering device based on LC-tunable Fabry-Perot (FP) nanocavities, demonstrating continuous control of the emission angle up to a maximum of 28°. Time-resolved photoluminescence (TRPL) measurements reveal, moreover, a 3-fold enhancement in cavity radiation emission due to its resonant character. Importantly, due to the coupling between TE and TM modes mediated by the LC birefringence, and the subsequent emergence of spin-orbit coupling, we further access control over the PL polarization state alongside the deflection angle. This approach shows promise in improving integration of dynamically-tunable emissive devices, addressing some of the challenges in miniaturization of this type of systems.

## Results
### Device design and fabrication

The proposed device is schematically depicted in Figure 1a. It comprises an active gain medium and a tunable material sandwiched between two distributed Bragg reflectors (DBR), forming a Fabry-Pérot cavity. The device further comprises an aluminum (Al) electrode beneath the bottom DBR, which also acts as a mirror, and a top transparent one above the top DBR, made of indium tin oxide (ITO). For the emitter, we use a quasi-2D perovskite (($PEA)_2FA_7Pb_8Br_{25}$), which was deposited as a thin layer onto the bottom DBR, while for the tunable material, we employ a commercial nematic liquid crystal (LC). A thin PMMA spacer layer is spin-coated on top of the perovskite layer, to physically separate it from the LC, preventing degradation of the gain[53]. For alignment purposes, we deposited a thin layer of polyimide on the top DBR and mechanically rubbed it, so that it pre-aligns the LC along the x-axis. The LC was then filled between the top and bottom DBRs after the cell is formed using capillary forces (see Methods for details). In this configuration, when a voltage is applied between the Al electrode and the ITO one, the liquid crystal rotates within the x-z plane (we define its angle of rotation, $θ_{LC}$, with respect to the z-axis, so that it is 0° when the LC is fully vertical). The extraordinary and ordinary refractive indices of the LC used (QYPDLC-001C) are $n_e$ = 1.813 and $n_o$ = 1.522, respectively.

The DBRs are composed of alternating layers of $TiO_2$ and $SiO_2$, optimized for high reflectivity across the visible spectrum (see Supplementary Note 1). These DBRs enable FP resonances that effectively confine light within the cavity, enhancing the optical feedback and supporting the resonant modes necessary for device operation. Importantly, the top DBR was designed with slightly lower reflectivity for wavelengths below 500 nm to facilitate the penetration of pump light into the device (pump = 488 nm, PL at 542 nm; see Supplementary Fig. 4). The cross-sectional scanning electron microscopy (SEM) images of the top and bottom DBRs are shown in Figures 1b and 1c. The optimized device structure exhibits smooth and uniform interfaces between each layer.

As the applied voltage increases, the director angle of the LC changes, leading to a change in the refractive index and, consequently, a shift in the resonant frequencies of the FP cavity. The simulated reflection spectra at different LC rotation angles are shown in Fig. 1d for both Transverse Electric (TE, polarized along the x-axis) and Transverse Magnetic (TM, polarized the y-axis) modes. As expected, TM modes are insensitive to the LC rotation, as they always experience the ordinary index of the LC ($n_o$), while TE modes show a significant blueshift as the rotation angle decreases (i.e. as the molecules are more vertically oriented) and the LC refractive index experienced by the TE wave transitions from $n_e$ to $n_o$. As a result, TE and TM modes can cross each other, something that, in our case, happens at an LC rotation angle of around 0.77 radians (~44 degrees). The corresponding experimental results are shown in Fig. 1e. Therein, a voltage ranging from 0 to 5 $V_{rms}$ at a frequency of 1 kHz is applied to the electrodes (all along this work, when talking about voltages, we refer to Voltage Root Mean Square). The device exhibits a switching threshold around 1$V_{rms}$, above which the TE modes undergo the expected blueshift while the TM modes remain unchanged, leading to a crossing at ~2.2 $V_{rms}$ of applied voltage.

**Dynamic photoluminescence tuning**

We measured the time-resolved photoluminescence both inside and outside the cavity, as shown in Fig. 2a, and fitted the data using a bi-exponential function[54-56] (see Supplementary Note 2). Outside the cavity, the slow component of the TRPL decay time is 450 ns, whereas inside the cavity, the slow component is reduced to 142 ns, indicating an enhancement factor of 3.16 within the cavity (see Supplementary Fig. S3) suggesting that the emission is indeed efficiently coupled to the cavity modes[57].

In Figs. 2b-f, we show color maps with the angle-resolved (in the yz-plane) photoluminescence spectra measured at different applied voltages (see methods' section for measurement details). First, we note the spectral shaping and directionality of the observed PL (see Supplementary Note 2 and Supplementary Fig. 4 for the comparison with the PL on a bare quartz substrate), exhibiting a dispersion that closely follows that of standard FP modes. When no bias is applied (0 V), bright emission through a mode with a central wavelength of 543 nm (at 0 degrees), is observed. Upon applying voltage, this mode gradually blueshifts up to the point that, at 4.8 $V_{rms}$, it moves beyond the lower limit of the emission bandwidth of the perovskite, leading to its disappearance. Simultaneously, at 0V, one can observe a second mode with a central emission wavelength of 570 nm (considering again the position for vertical emission). Upon applying voltage, this mode blueshifts, becoming the mode with the highest emission intensity at 4.8 $V_{rms}$, with a peak ~540 nm. Finally, a mode that was initially beyond the emission bandwidth, and thus not observed at zero bias, first faintly appears at 2.4 $V_{rms}$, and then blueshifts for increasing bias, up to a final peak position of 575 nm when the voltage is increased to 4.8 $V_{rms}$. As evidenced from the figure, the peak emission of these modes in the normal direction and their behavior with increasing voltage matches the trends observed in the reflection spectrum. The continuous blue-shifting of the photoluminescence bands can be better visualized in the Supplementary Video 1. For completeness, we also performed the reverse voltage scanning, showing good consistency with the forward voltage scanning results (see Supplementary Note 4 and Supplementary Fig. 5).

**Optical Spin-orbit coupling**

Beyond the dynamic tuning of the PL emission wavelength, an interesting feature can be observed in the angle-resolved PL spectral maps of Fig. 2. Namely, after some particular voltage, the PL bands appear to split into two for non-zero emission angles (becoming more pronounced as the angle increases). For the two bands initially observed in the PL maps at zero bias, this splitting starts to occur for voltages ~2 $V_{rms}$. Interestingly, as the applied voltage increases, the splitting first increases and then, beyond a certain value, it gradually weakens, to the point that it fully disappears when 4.8 $V_{rms}$ is applied (a situation in which the LC is almost fully vertically rotated). This behavior can be explained by the birefringence-mediated coupling between the TE and TM modes supported by this system, which happens when modes of opposite polarization and parity cross each other, thereby being able to exchange energy[58-61]. Indeed, by looking again at Figs. 1d-e, one can see that this is exactly what happens in the voltage range in which splitting is observed. Therein, we observe a clear crossing between TE-TM modes for voltages ~2 $V_{rms}$. In such a scenario, one would expect to observe cavity-induced spin-orbit coupling (SOC), causing the emitted light to undergo a transformation from linearly polarized emission into circularly polarized one[58-61].

To confirm this, we measured the Stokes parameters of the PL[62] (see Methods), as shown in Fig. 3d-i. At 1 $V_{rms}$, below the mode crossing, the S1 parameter approaches unity in absolute value, while the S3 parameter values are low, indicating that the emitted light is primarily linearly polarized. On the contrary, at 2.4 $V_{rms}$ (i.e., when TE-TM modes cross each other and the splitting in PL is observed) the S1 parameter values become relatively low, particularly for angles >5° in absolute value, and the S3 parameters approache unity values. In this situation, the emitted light is primarily circularly polarized. We note a clear crossover between left- and right-handed circularly polarized light as one moves from one of the split branches of the mode to the other. At 4.8 $V_{rms}$, the S1 parameter becomes dominant again, indicating a return to predominantly linearly polarized emission. These observations are in very good agreement with the measured Stoke's parameters of the light reflected by the system at the emission wavelength (taken as 532nm). Indeed, we show these in the momentum space images in Supplementary Note 5 and Supplementary Fig. S6, where a clear splitting of the light cones in the $k_y$ direction (corresponding to the yz-plane of emission) is observed under circular polarization incidence of different handedness.

To corroborate the measured results, we first numerically calculate the Stokes parameters of the PL using full-wave numerical simulations for three different LC rotation values (see Fig. 3g-i), corresponding to situations before, at, and after the crossing point of the TE-TM modes, finding consistent results with the experimental results. To gain further insight into the physical origin of the observed polarization behavior, we refer to a Rashba–Dresselhaus spin-orbit coupling model in a planar DBR cavity. The form of the effective Hamiltonian and analysis of its eigenvalues and eigenstates, is discussed in detail in in the Supplementary Note 6.

**Dynamic PL beam steering**

We now use the PL coupling into the nanocavity modes and the resonant mode shift induced upon biasing to dynamically change the PL emission angle at specific wavelengths. To do so, we consider the situation in which a bandpass filter is included to filter the wavelength of interest. We mimic this through backend programming to selectively isolate and process signals within the target wavelength range.

Figure 4a presents the angle-resolved PL within a chosen band (543-546 nm) for different applied voltages. As can be seen, a main mode (labelled as *mode1*) can be observed, whose angle of emission can be widely tuned, ranging from normal emission up to almost ±28°. A secondary mode, with lower tunability in the applied voltage range, is also present (labelled as *mode2*). In Fig. 4b, we plot the beam deflection angles for these two modes as the voltage varies from 0 $V_{rms}$ to 5 $V_{rms}$, showing that indeed the central emission angle (*mode1*) can be continuously tuned from ±28° to 0°. (see Supplementary Video 2)

**Dynamic polarization conversion**

Next, we measure the Stokes parameters (S1 and S3) of the dynamically steered PL bands, as shown in Figure 4c-f. As can be seen there, the main PL steering mode (*mode1*) retains an almost-linear polarization character (S1~-1) from normal emission up to ±16°. At that point, the PL becomes predominantly circularly polarized and quickly changes back to linear polarization (but with opposite polarization, S1~1). Interestingly, beyond that switching point and for larger angles, it is *mode2* that becomes predominantly linearly polarized with S1~-1, indicating that if a linear polarizer (S1~-1) would be used, one could continuously tune the PL angle in an even wider range than that provided by *mode1* alone. To further demonstrate the versatility of the functionality, we now make use of the above-mentioned crossing point, in which the polarization changes from linear to circular and back, to enable polarization conversion at a specific wavelength and emission angle. To better illustrate this possibility, the polarization states corresponding to the PL at different voltages and for two angular ranges are marked on the Poincaré sphere[63,64] in Figs. 5a and 5b, where S1, S2, and S3 are the spherical coordinate axes.

In the emission angle range from -25° to -27° (Fig. 5a), as the applied voltage increases from 1V to 2.6V, the polarization state initially positioned near the equator of the Poincaré sphere gradually transitions towards the vicinity of the S3 pole, indicating an evolution from linear polarization to nearly circular polarization.

In the emission angle range from -27° to -29°, as the voltage is increased from 2.6 $V_{rms}$ to 5 $V_{rms}$, the polarization state initially located between the S2 and S3 axes on the Poincaré sphere gradually shifts towards the region between the S1 and S3 axes, indicating a transition from one form of elliptical polarization to another, with changes in orientation.

Projections of the electric field trajectory in the xy-plane are shown in Figs. 5c and 5d, corresponding to the different Polarization conversions presented in Fig. 5a and 5b, showing that a larger degree of dynamic control over the PL polarization is indeed possible.

**Conclusion**

In conclusion, we have demonstrated a larger dynamic control over the photoluminescence emitted from a perovskite coupled to a liquid crystal-based, tunable Fabry-Perot nanocavity. Using it, we have shown dynamic PL steering, achieving a maximum deflection angle of 28° with continuous beam steering capabilities. Based on the spin-orbit coupling effect, and the integration of liquid crystals as a tunable birefringent medium, we also access precise control over the polarization state of the emitted light, facilitating both polarization gradient creation and polarization conversion at specific emission angles. Although not specifically discussed here in details, this type of devices can have switching speeds on the millisecond time scales (see Supplementary Note 5 and Supplementary Fig. 7), or even faster if properly optimized.

The findings represent a significant advancement in the field of beam steering technologies, offering a promising, fully integrated solution to overcome current limitations in angle tunability and emission directivity. The potential for high system integration and dynamic control makes this device a compelling candidate for a wide range of applications, including telecommunications, sensing and imaging, paving the way for next-generation optical devices with enhanced performance and versatility.

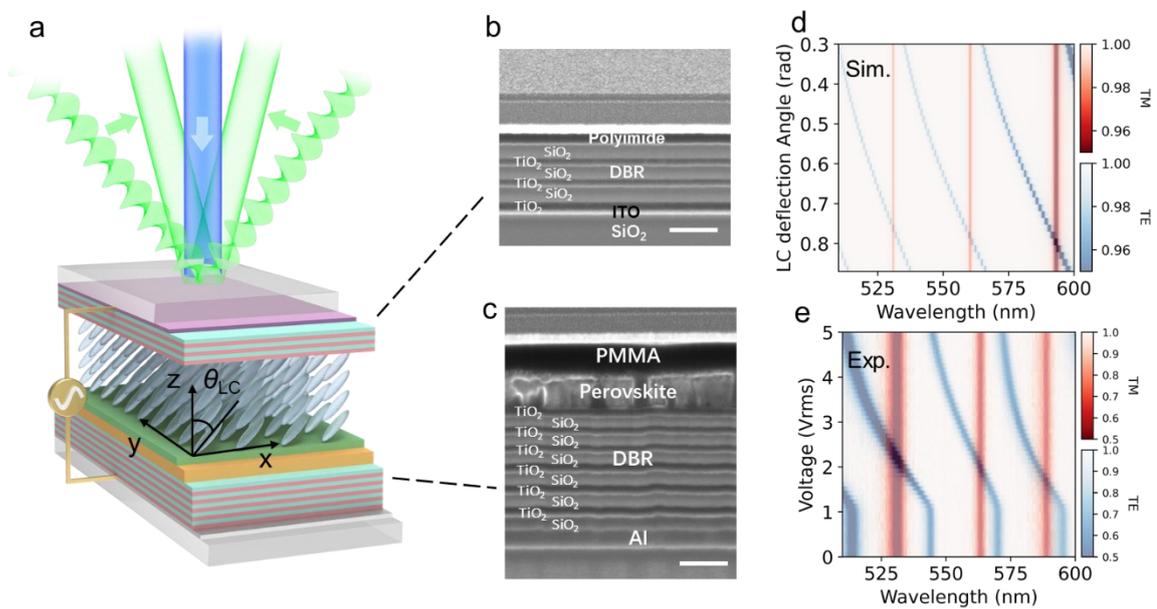

**Fig 1. Schematic description and characterization of the proposed device. a** schematic of the device **b** cross section of top DBR and **c** bottom DBR with quasi 2D perovskite, shown in scanning electron microscope images. The scale bar is 250 nm. **d** Simulated and **e** experimental reflectance spectra under normal incidence as functions of LC deflection angle ($\theta_{LC}$) and applied bias voltage ($V_{rms}$), respectively. The TE mode blue shifts with the voltage and intersects with the TM mode.

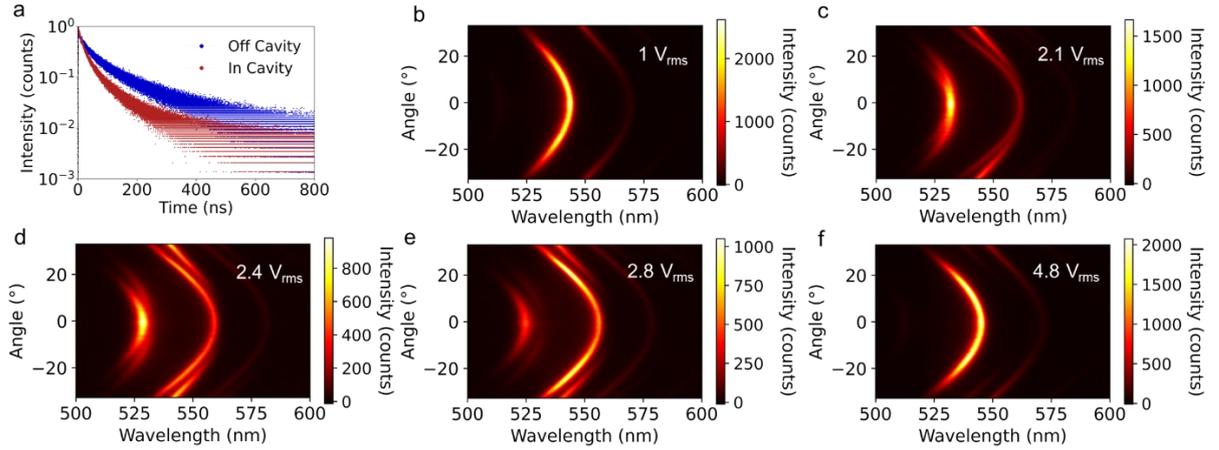

**Fig 2. Electrically tunable Photoluminescence. a** Time resolved photoluminescence of quasi 2D perovskite outside and inside the FP cavity. **b-f** Measured angle-resolved photoluminescence under different voltages (1 $V_{rms}$, 2.1 $V_{rms}$, 2.4 $V_{rms}$, 2.8 $V_{rms}$ and 4.8 $V_{rms}$). The colormap indicates the value of PL counts. As the voltage increases, the PL modes exhibit a noticeable blue shift and, for some voltage values, a clear splitting at off-normal emission angles.

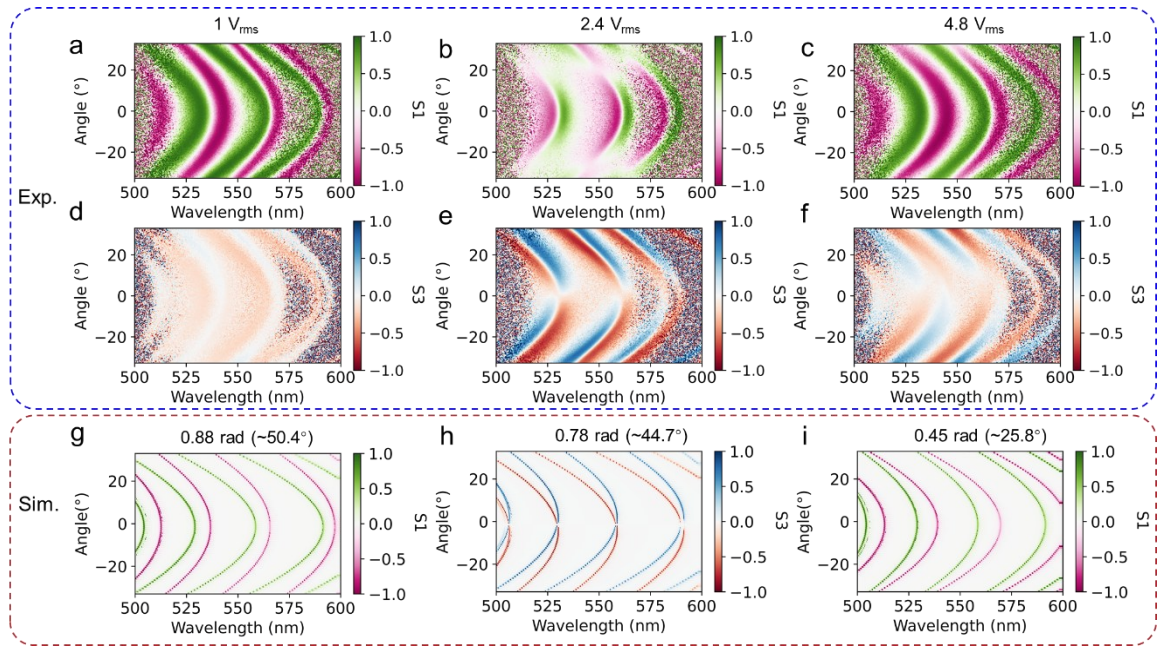

**Fig 3. Demonstration of PL polarization transformation. a-f.** Angle-resolved S1 and S3 parameters measured at 1V (a, d), 2.4V (b, e) and 4.8V (c, f). The device is operating before and after the TE-TM mode crossing in (a, d), and (c, f), respectively. The device is operating at the mode crossing in (b, e), where coupling of TE and TM modes with opposite parity transforms the emitted PL polarization from linear to circular. **g-i.** Simulated angle-resolved S1 (g, i) and S3 (h) parameter at LC rotation angles of $\theta_{LC}$ = 0.88 rad (g), $\theta_{LC}$ = 0.45 rad (i) and $\theta_{LC}$ = 0.78 rad (h). The S3 parameters for $\theta_{LC}$ = 0.88 rad and $\theta_{LC}$ = 0.45 rad and the S1 parameters for $\theta_{LC}$ = 0.78 rad are negligible and not shown here.

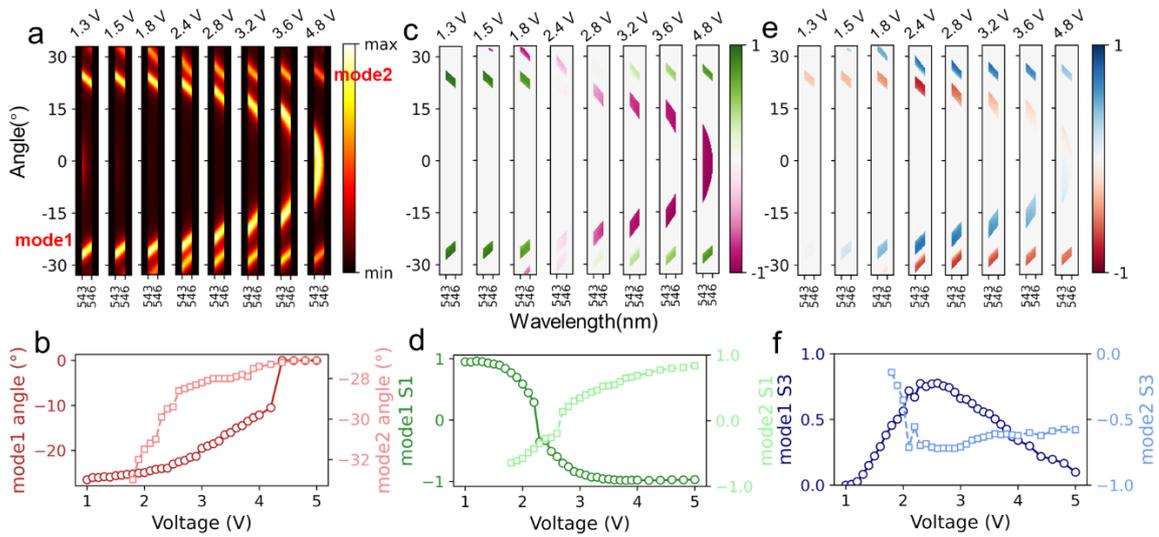

**Fig 4. Dynamic beam steering with polarization analysis. a** Photoluminescence with a bandpass filter (543–546 nm) under different voltages; from left to right, the applied voltages are 1.3 V, 1.5 V, 1.8 V, 2.4 V, 2.8 V, 3.2 V, 3.6 V, and 4.8 V. **b** Emission angle variation of the two optical modes as a function of applied voltage (1 V to 5 V). **c** The Stokes parameter S1 measured under the same experimental and voltage conditions as in (a). **d** Voltage-dependent evolution of the S1 parameter for the two modes, measured from 1 V to 5 V. **e** The Stokes parameter S3 measured under the same experimental and voltage conditions as in (a). **f** Voltage-dependent evolution of the S3 parameter for the two modes, measured from 1 V to 5 V.

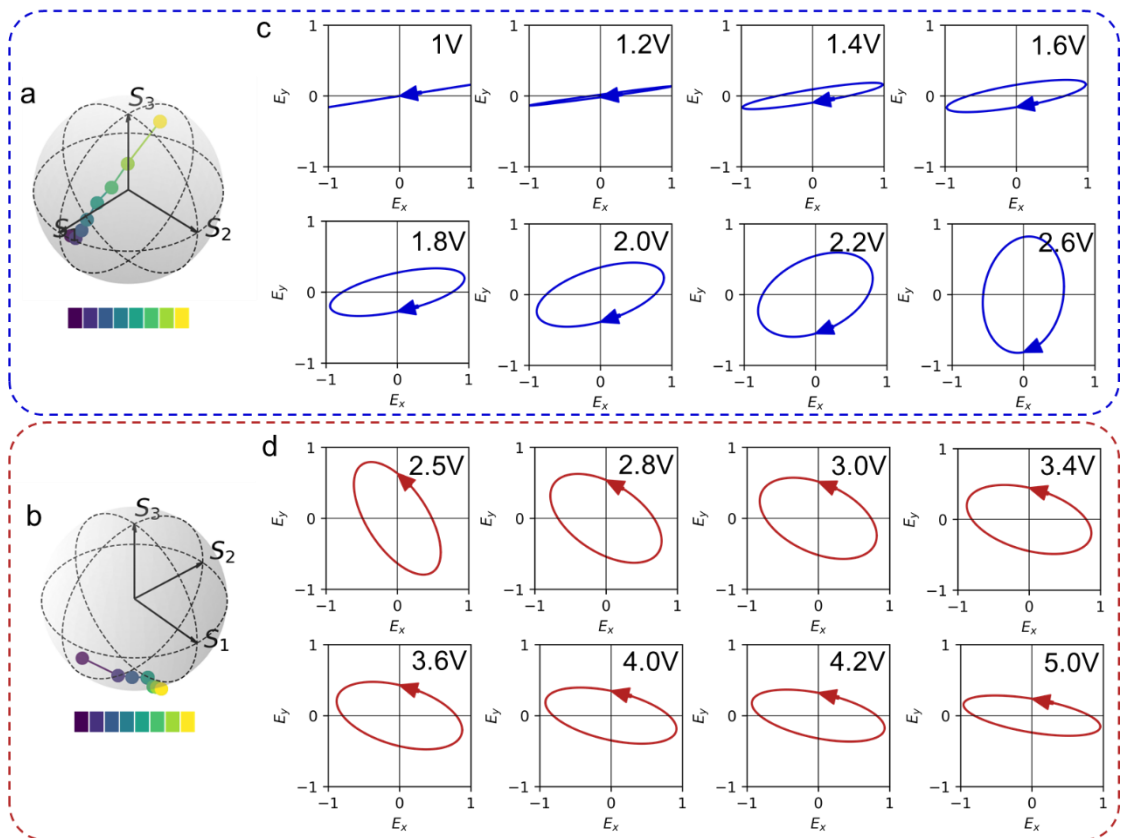

**Fig 5. Polarization conversion at selected emission angles. a, b** Distributions of polarization under different voltages are displayed on the Poincaré sphere at emission angle of -25° to -27°(a) and -27° to -29°(b). **c, d** Projections of electric field trajectory in the xy-plane under different voltages at emission angles of -25° to -27°(c), drawn with blue lines, and -27° to -29°(d), drawn with red lines. Each figure is labeled with the corresponding voltage value.

**Methods**

Simulation methodology

The far-field emission pattern was calculated using reciprocity based on a commercially available, full wave electromagnetic solver (COMSOL Multiphysics). For the simulation of the Stokes parameters in Fig.3, we use TE and TM as incident light for S1 and LCP and RCP for S3. A single unit cell was simulated with Bloch boundary conditions applied laterally to replicate an infinite system. Both the excitation and receiving ports were periodic.

Device fabrication

The fabrication of the device begins by depositing a 150 nm thick aluminum (Al) layer onto an 8-inch silicon wafer, which features a 1 μm thick thermal silicon dioxide ($SiO_2$) layer, using a physical vapor deposition (PVD) system (AMAT Endura). The wafer is then diced into substrates of 20 mm × 20 mm. The Al layer is used to form the bottom electrodes. The bottom distributed Bragg reflector (DBR) is formed from 12 alternating layers of TiO2 and SiO2, where the $TiO_2$ and $SiO_2$ are deposited using an ion-assisted deposition (IAD) tool (Oxford Optofab3000). The perovskite layer and the PMMA layer are then successively spin-coated onto the bottom DBR.

The top DBR structure, consisting of six alternating $TiO_2$ and $SiO_2$ layers, is deposited similarly, but in this case in an indium tin oxide (ITO) coated (23 nm thick) glass substrate. A polyimide layer is spin-coated onto the top DBR, then cured at 150°C for 30 minutes on a hot plate. The cured substrates are unidirectionally rubbed with soft velvet cloths to establish a preferred alignment direction for the liquid crystal layer. The rubbing is performed using a commercially available rubbing machine (Holmarc Opto-Mechatronics - HO-IAD-BTR-03), with controlled rubbing force. Nematic liquid crystal QYPDLC-001C is then introduced into the cell via capillary action. To verify the uniform alignment of the liquid crystal, the device is observed under a microscope with crossed polarizers.

Perovskite：

The perovskite is deposited in a nitrogen glovebox The precursor solution for $(PEA)_2FA_7Pb_8Br_{25}$ was prepared by dissolving phenylethylammonium bromide (PEABr), formamidinium bromide (FABr), and lead bromide ($PbBr_2$) at a molar ratio of 1:4:4 in DMF with a $Pb^{2+}$ concentration of 0.4 M. The mixture was stirred at 60 °C for 12 hours. 200 μl of the resulting solution was spin-coated onto the bottom DBR at 3,000 r.p.m. for 60 seconds. The substrates were then heated on a hotplate at 70 °C for 15 minutes, followed by 100 °C for 5 minutes.

A 5 wt% solution of PMMA in anisole was used as the spacer layer material. Anisole was chosen as the solvent because of lower polarity compared to the perovskite precursor solvent DMF. This lower polarity prevents the disruption or dissolution of the perovskite film when the PMMA solution is applied. The PMMA solution was spin-coated onto the perovskite film at a speed of 4000 r.p.m., forming a uniform and compact polymer spacer layer.

Measurement:
Angle-resolved spectroscopic measurements were performed using an inverted optical microscope (Nikon Ti-U) coupled to a spectrometer equipped with an electron-multiplying charged-coupled detector (EMCCD, Andor Newton 971). For photoluminescence measurements, the excitation was a 488 nm continuous-wave laser. The signals were collected by a 50× objective (Nikon, NA = 0.55) and passed through a series of lenses for back focal plane imaging. The wavelengths were resolved by a spectrograph (Andor SR-303i) with a single grating groove density of 150 gr/mm and a slit size of 100 μm. The angle was scanned along the y-direction.

For the reflection and PL measurements of Figs. 1 and 2, respectively, a linear polarizer was set at the collection. For the Stokes parameters measurement, a linear polarizer, a half-wave plate and a quarter-wave plate were placed in the collection path. The Stokes parameters were calculated from the PL measurements in various polarization states: linear polarization with polarization plane oriented in horizontal ($I_H$), vertical ($I_V$), diagonal ($I_D$), antidiagonal ($I_A$) directions and both circular polarizations ($I_{\sigma^+}$, $I_{\sigma^-}$), from which they are extract using the formulas:

$$S_1 = \frac{I_V - I_H}{I_V + I_H}, \ S_2 = \frac{I_D - I_A}{I_D + I_A}, \ S_3 = \frac{I_{\sigma^+} - I_{\sigma^-}}{I_{\sigma^+} + I_{\sigma^-}}$$

Note that, in this notation, the TE mode corresponds to the horizontal polarization, while the TM mode corresponds to the vertical polarization.

## Acknowledgements
The authors acknowledge the funding support from the Singapore MTC Programmatic Grant No: M21J9b0085 and the Frontier Technology Research and Development Program of Jiangsu Province, China (No: BF0408). C. Dong thanks the support from the China Scholarship Council.

Supplementary information for "Dynamic Control of Momentum-Polarization Photoluminescence States with Liquid-Crystal-tuned Nanocavities"

**Chengkun Dong, Matthew R. Chua, Rasna Maruthiyodan Veetil, T. Thu Ha Do, Lu Ding, Deepak K. Sharma, Jun Xia, Ramón Paniagua-Domínguez***

*Email: ramon_paniagua@imre.a-star.edu.sg

## Supplementary Note 1: Design and characterization of bottom and top DBRs

The thickness of each dielectric layer in the DBR is adjusted to ensure a good reflectivity across the visible spectrum. For that, we use six $SiO_2/TiO_2$ pairs for the bottom DBR, comprising three pairs optimized for peak reflectivity at 450nm and three pairs for peak reflectivity at 530nm. This gives a good overall reflectivity across the visible, and results into the following thicknesses for each layer:

$$Thickness\_1d\_SiO_2 = \frac{\lambda_{1d}}{4n_{SiO_2}} = \frac{450nm}{4 \times 1.46} = 77.05nm$$

$$Thickness\_1d\_TiO_2 = \frac{\lambda_{1d}}{4n_{TiO_2}} = \frac{450nm}{4 \times 2.48} = 45.36nm$$

$$Thickness\_2d\_SiO_2 = \frac{\lambda_{2d}}{4n_{SiO_2}} = \frac{530nm}{4 \times 1.46} = 90.75nm$$

$$Thickness\_2d\_TiO_2 = \frac{\lambda_{2d}}{4n_{TiO_2}} = \frac{530nm}{4 \times 2.48} = 53.42nm$$

where $n_{layer}$ is the refractive index of the layer and $\lambda_j$ the optimization wavelength.

The upper DBR comprises three pairs, with one pair optimized for peak reflectivity at 580nm and two pairs for peak reflectivity at 450nm. The resulting layer thickness are as follows:

$$Thickness\_1u\_SiO_2 = \frac{\lambda_{1u}}{4n_{SiO_2}} = \frac{500nm}{4 \times 1.46} = 85.62nm$$

$$Thickness\_1u\_TiO_2 = \frac{\lambda_{1u}}{4n_{TiO_2}} = \frac{500nm}{4 \times 2.48} = 50.40nm$$

$$Thickness\_2u\_SiO_2 = \frac{\lambda_{2u}}{4n_{SiO_2}} = \frac{580nm}{4 \times 1.46} = 90.75nm$$

$$Thickness\_2u\_TiO_2 = \frac{\lambda_{2u}}{4n_{TiO_2}} = \frac{580nm}{4 \times 2.48} = 53.42nm$$

Supplementary Figures 1 and 2 show schematics of the DBRs, together with simulated and measured reflectivity values.

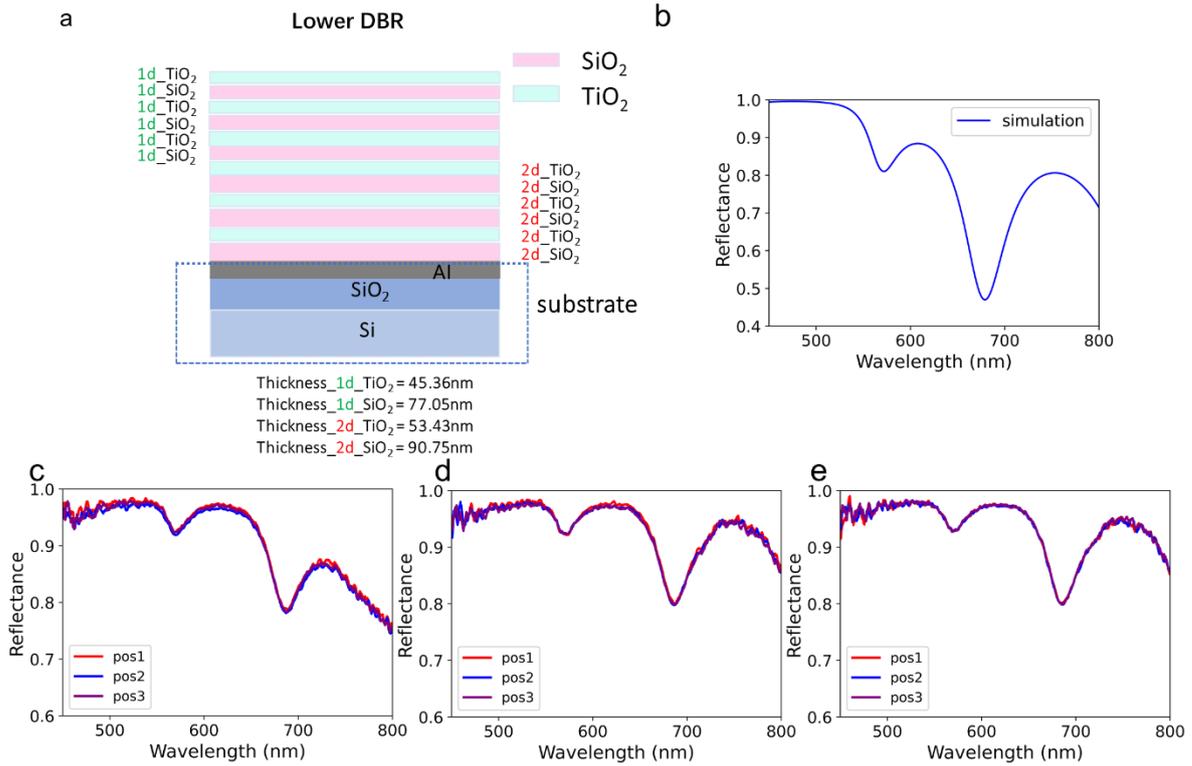

Supplementary Fig. 1: **a,** The schematic of lower DBR. **b,** is the simulated reflection spectrum of 450nm-800nm. **c-e,** are the measured reflectance spectra of three samples, and three positions are randomly selected on each sample.

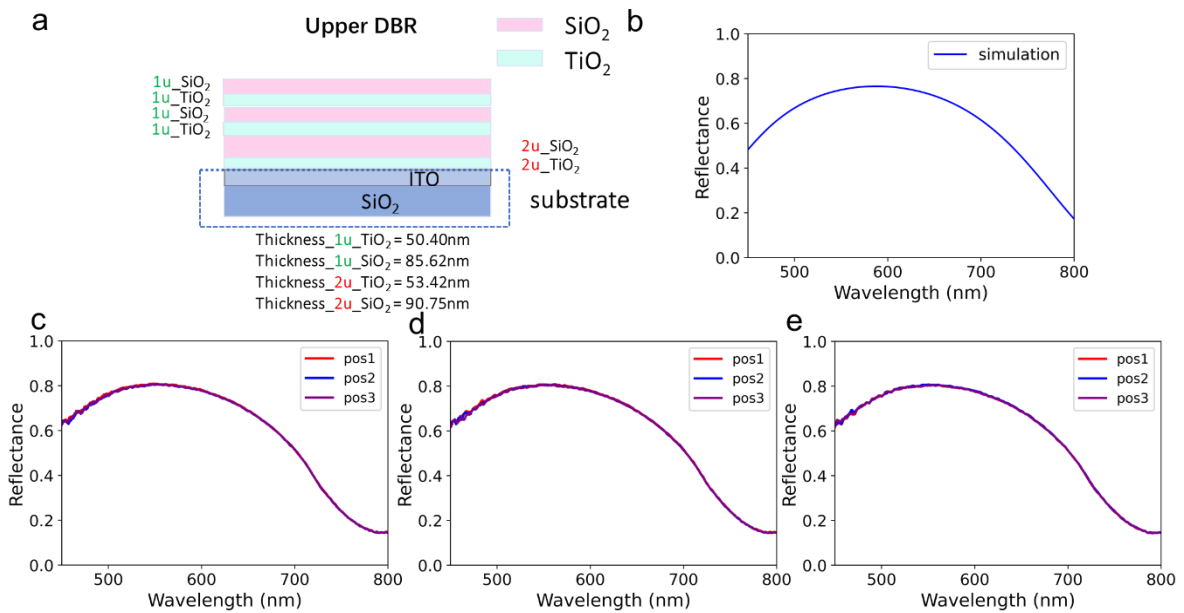

Supplementary Fig. 2: **a,** shows the schematic of top DBR. **b,** is the simulated reflection spectrum of 450nm-800nm. **c-e,** are the measured reflectance spectra of three samples, and three positions are randomly selected on each sample.

## Supplementary Note 2: Lifetime measurements and fitting

The **decay rate ($\Gamma$)** is defined as the power ($P$) dissipated by the dipole emitter divided by the photon energy ($\hbar\omega$). The average dissipated power is expressed mathematically as:

$$\Gamma = \frac{P}{\hbar\omega}$$

$$P = \langle\frac{dW}{dt}\rangle = \frac{\omega^3}{2c^2\varepsilon}Im\{\boldsymbol{p}^*\bar{\bar{G}}_{tot}(r0,r0,\omega)\boldsymbol{p}\}$$

here **$P$** represents the dipole moment, $\bar{\bar{G}}_{tot}$ is the Green's function describing the photonic environment, $r0$ is the emitter's position, and $\omega$ is the angular frequency

The Purcell factor quantifies the enhancement of an emitter's radiative decay rate caused by its surrounding photonic environment. Only a portion of the dissipated power contributes to radiation, with the remaining energy being absorbed or transformed into heat within the photonic environment. The radiative power ($P_{rad}$) can be calculated as:

$$P_{rad} = \Gamma_{rad} \times \hbar\omega$$

where $\Gamma_{rad}$ is the radiative decay rate. The total decay rate is then expressed as the sum of the radiative ($\Gamma_{rad}$) and non-radiative ($\Gamma_{non-rad}$) components:

$$\Gamma = \Gamma_{rad} + \Gamma_{non-rad}$$

The photoluminescence decay curve of perovskite maiterals can be fitted with a bi-exponential function:

$$I(t) = A_1 e^{-t/\tau_1} + A_2 e^{-t/\tau_2}$$

$A_1$ and $A_2$ are the amplitudes of the two exponential components, indicating the relative contributions of each decay process. $\tau_1$ and $\tau_2$ are the time constants associated with each exponential decay, representing the characteristic decay times of the two processes.

Typically, at low excitation fluence, the shorter decay rate is attributed to trap-assisted recombination, while the longer lifetime component is attributable to the radiative decay process.

In our example, fitting results show that the slow component of the perovskite material outside the FP cavity is 450 ns, while inside the FP cavity, the slow component is 142 ns. Therefore, the Purcell factor is enhanced by 3.17 time

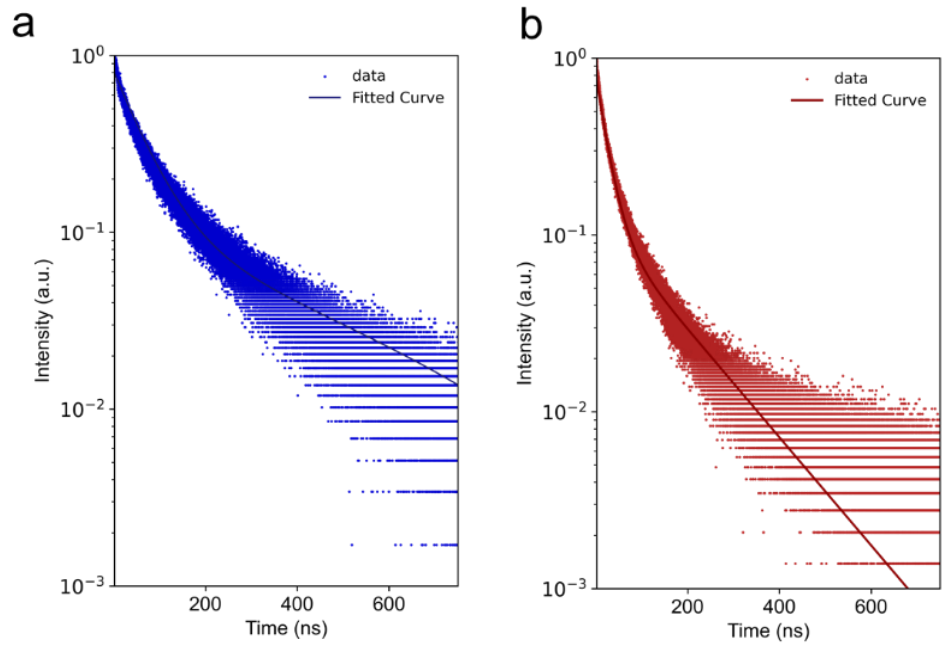

|  | Slow component |
|---|---|
| outside cavity | 450 ns |
| inside cavity | 142 ns |

Supplementary Fig. 3: time-resolved photoluminescence **a,** outside and **b,** inside cavity. The data were fitted using a bi-exponential decay model, revealing faster decay dynamics in the in-cavity case due to enhanced spontaneous emission facilitated by the cavity resonance.

## Supplementary Note 3: Comparison of PL on quartz and FP cavity

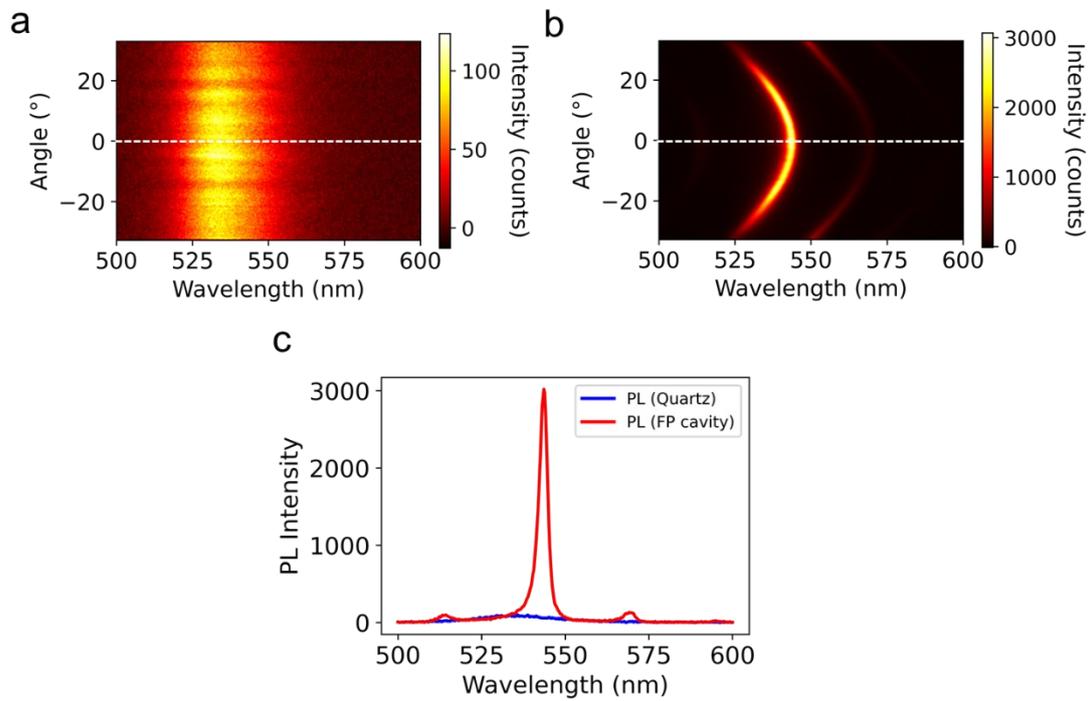

Supplementary Fig. 4: Comparison of PL on quartz and FP cavity. **a,** angle-resolved PL on quartz **b,** angle-resolved PL in FP cavity, illustrating the enhanced angular confinement and directionality of the emitted light due to the cavity's resonant modes. **c,** Comparison of PL intensity in two cases under normal emission (white dashed lines in **a** and **b**)

# Supplementary Note 4: Reflection and PL spectra in the normal direction upon forward and reverse voltage sweep

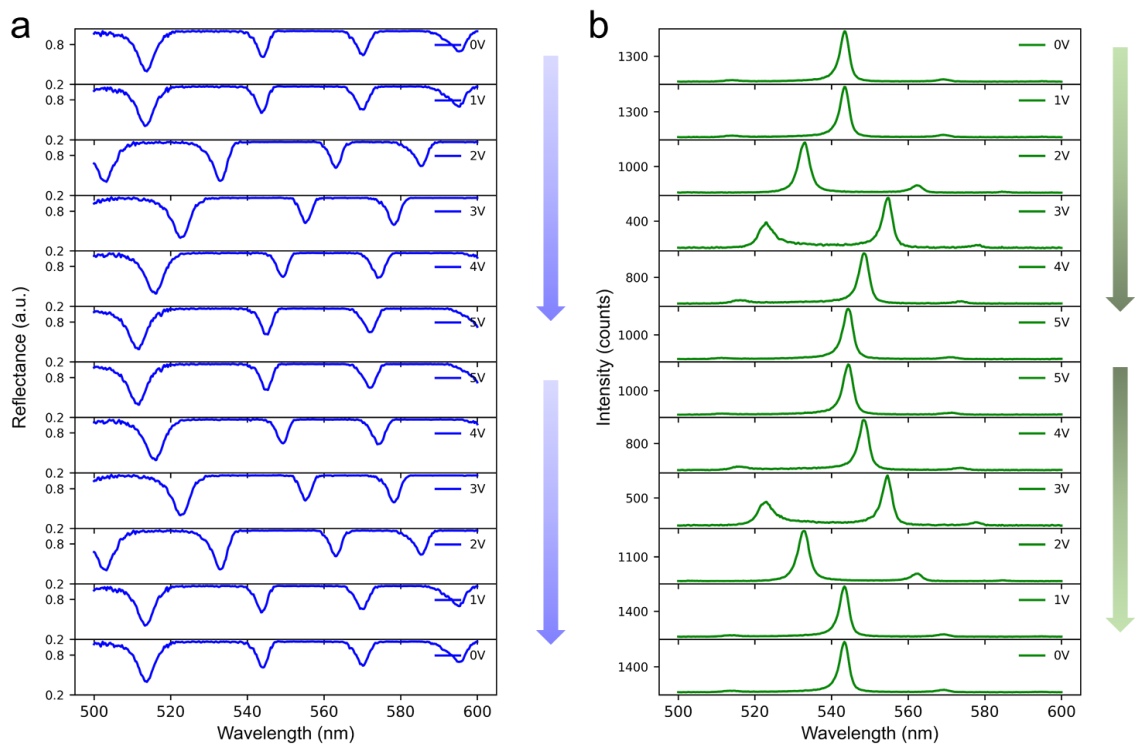

Supplementary Fig. 5: **a,** Reflectance spectra measured at normal incidence during forward (1V to 5V) and reverse (5V to 1V) voltage sweeps. **b,** Photoluminescence spectra (normal emission) for forward and reverse voltage sweeps, demonstrating repeatability of the optical response. We applied a voltage sweep from 1V to 5V, followed by a reverse sweep from 5V back to 1V. The experimental results revealed a high degree of consistency between the forward and reverse voltage sweeps[1], demonstrating the stability and repeatability of the system optical response under cyclic voltage modulation. Notably, the absence of any memory effect further underscores the robustness and reliability of the sample's optical and electrical properties, making it highly suitable for applications requiring precise and reversible control.

# Supplementary Note 5: Polarization resolved back-focal plane measurements of the reflection at 532nm.

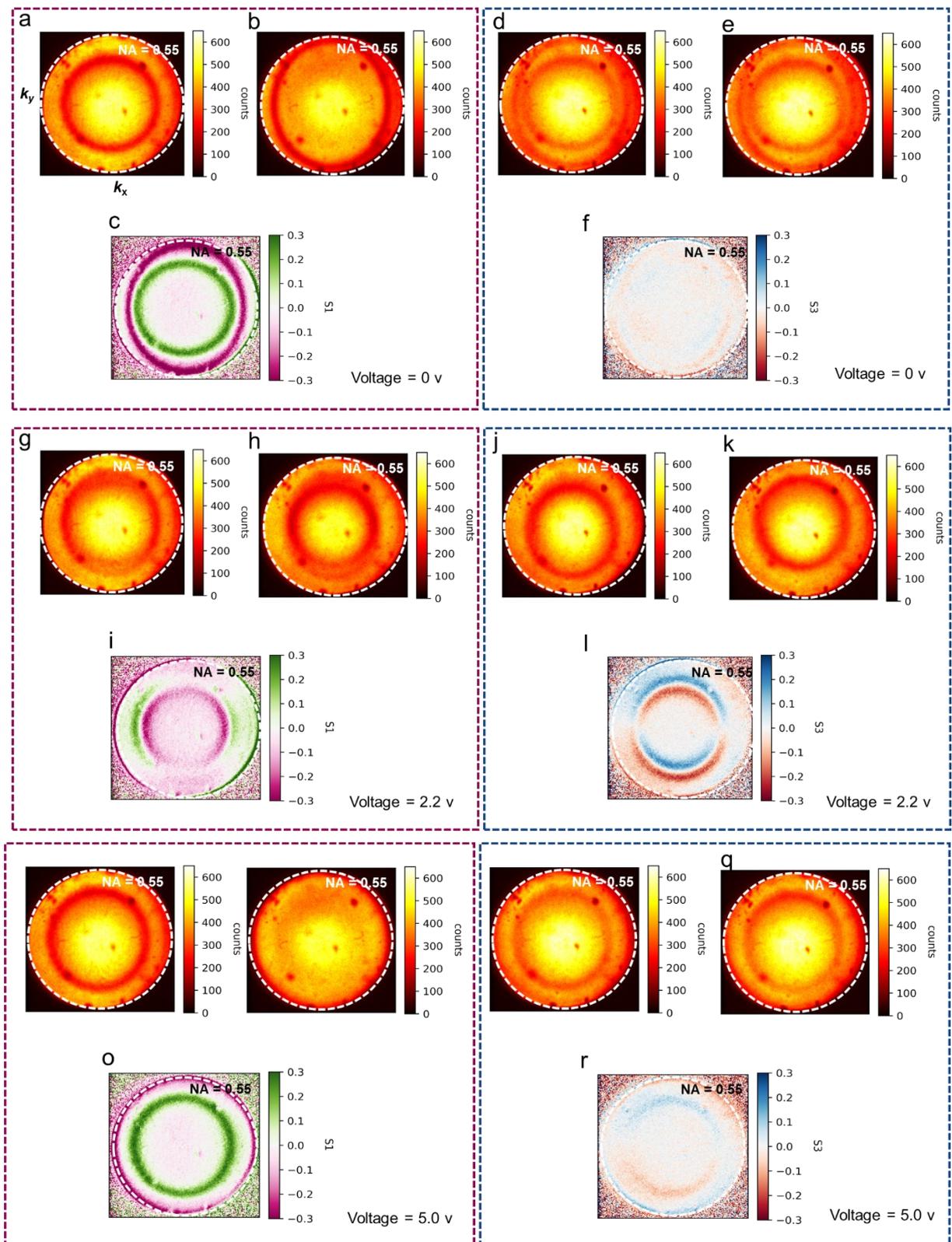

Supplementary Fig. 6: Back focal plane images of the device reflection at 532 nm, collected using a microscope objective with an NA = 0.55. These images correspond to three

conditions: **a-f,** before TE-TM mode crossing, **g-l,** at TE-TM mode crossing, and **m-r,** after TE-TM mode crossing. **a, g, m,** Reflection under vertical polarization. **b**, **h**, **n**, Reflection under horizontal polarization. **d, j, p,** Reflection under σ+ (left) polarization. **e, k, q,** Reflection under σ- (right) polarization. **c, i, o,** Calculated S1 parameter. **f, l, r,** Calculated S3 parameter. Upon TE-TM crossing, two offset spin-split circles in momentum space become apparent.

## Supplementary Note 5: Eigenvalue Evolution and Polarization Properties in Momentum Space.

To shed some light on the physical origin of the effect, we use a description based on the Rashba-Dresselhaus effect[2], which can be theoretically described by an effective 2×2 Hamiltonian in the circular polarization basis as follows:

$$H = \frac{\hbar k_x^2}{2m_x} + \frac{\hbar k_y^2}{2m_y} - 2\alpha \hat{\sigma}_z k_y + (E_X - E_Y)\hat{\sigma}_x,$$

Where $\hbar$ is the reduced Planck constant, $\alpha$ denotes the SOC coupling strength, $m_x$ and $m_y$ are the effective masses of the cavity photon in the xy plane, and $\hat{\sigma}_x$ and $\hat{\sigma}_z$ are the first and third Pauli matrices. In this model, as the LC rotates, the term $(E_X - E_Y)$ will change. When the TE and TM modes coincide at k=0, $(E_X - E_Y)$ becomes zero, resulting in two orthogonal linearly polarized modes, which then split into circularly polarized modes as the angle changes.

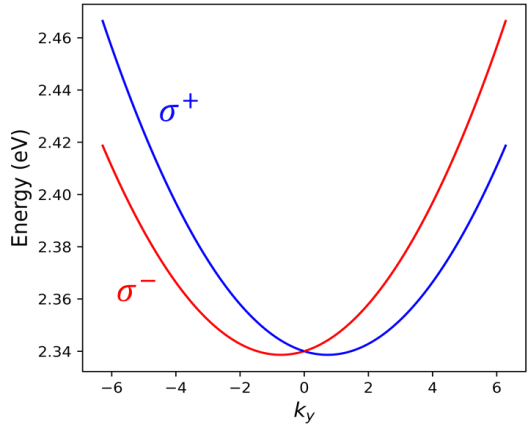

Supplementary Fig. 7. Energy eigenvalues of the model Hamiltonian as a function of ky. The eigenstates of the red curve exhibit left-handed circular polarization, whereas those of the blue curve exhibit right-handed circular polarization, as determined by the computed degree of circular polarization.

The figure depicts the variation of two eigenvalues (energy bands) as a function of ky at kx=0. By analyzing the corresponding eigenstates, the degree of circular polarization is calculated. In this model, the eigenstates of the two energy bands exhibit distinct polarization characteristics: the eigenstate corresponding to the red curve exhibits left-handed circular polarization, while that of the blue curve exhibits right-handed circular polarization.

## Supplementary Note 6: Response time measurement

The setup used for measuring liquid crystal response time follows a standard cross-polarizer configuration. In this system, the sample is oriented such that its liquid crystal director forms a 45° angle with the polarization of the incident light[3]. A 540 nm linearly polarized laser, produced by a supercontinuum light source with a Multi-Line AOTF filter (SuperK EXTREME and SuperK Select, NKT Photonics), is directed towards the sample. The direction of the analyzer is perpendicular to the polarization direction of the incident laser. A fast voltage pulse or modulation signal is used to excite the liquid crystal device, and the change in output light intensity over time is recorded. The response time is typically defined as the time it takes for the light intensity to reach 90% (or 10%) of its maximum change. The light intensity is detected by a photodiode (Thorlabs, PDA36A2 amplified detector), and the signal generator is coupled with the photodiode to the oscilloscope (LeCroy WavePro 715Zi-A 1.5GHz) to synchronize the excitation and response recordings.

We measured the response time of the device under 5 $V_{pp}$ and conducted a detailed analysis of its rise time and fall time. A signal with a frequency of 10 kHz was generated using a waveform generator to drive the device, allowing us to observe its dynamic response characteristics. Supplementary Fig. 8a records the changes in light intensity signal over multiple "on" and "off" cycles of voltage pulses. Supplementary Fig. 8b shows the "switch off" response, with the corresponding response time of 3 ms. Supplementary Fig. 8c records the "switch on" response, with the corresponding response time of 1.8 ms.

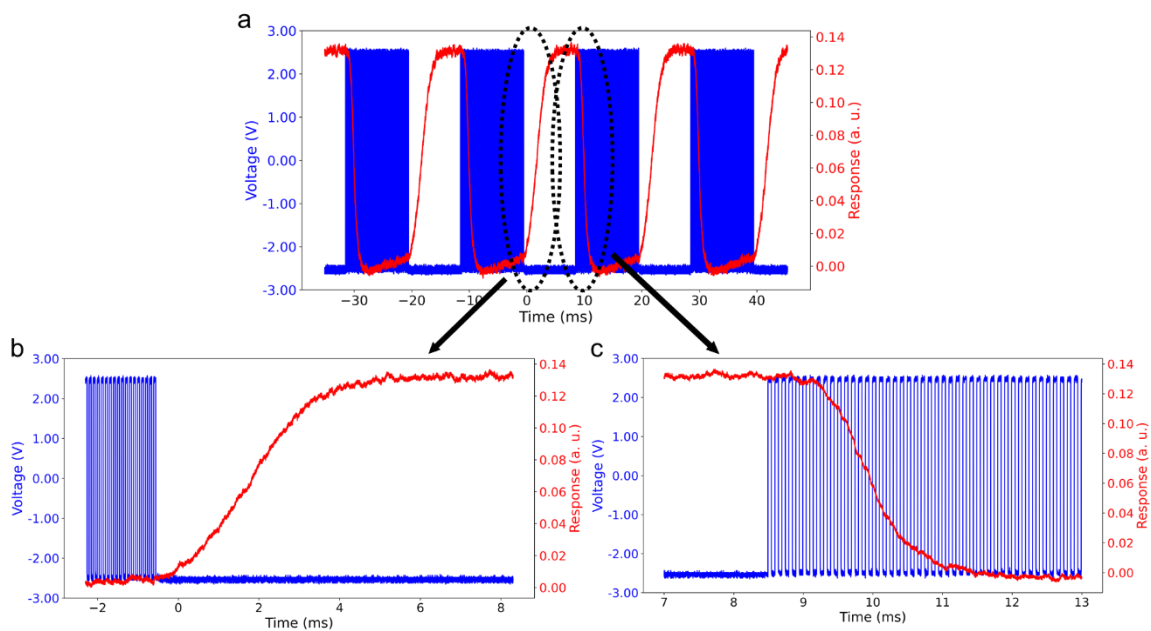

Supplementary Fig. 8: response time measurement. **a,** Multiple measurement cycles are repeated. **b,** Switching-off process, extracted from one measurement cycle, represents the time taken for the system to transition from its on state to the off state. **c,** Switch-on process, extracted from the same measurement cycle, represents the time required for the system to transition from the off state to the on state.

# Supplementary References